\documentclass[twocolumn,prb,aps,showpacs]{revtex4}

\usepackage{amsmath,amssymb}
\usepackage{pstricks}
\usepackage{epsfig}

\def\nind{\noindent}
\def\<{\left<}
\def\>{\right>}
\def\ket|#1>{\left|#1\right>}
\def\bra<#1|{\left<#1\right|}
\def\elem<#1|#2|#3>{\left<#1\right|#2\left|#3\right>}
\def\({\left(}
\def\){\right)}

\begin{document}

\title{Quantum Wavefunction Annealing of Spin Glasses on Ladders}

\author{J. Rodr\'{\i}guez-Laguna}
\affiliation{International School for Advanced Studies (SISSA), Via Beirut 2-4,
I-34014 Trieste, Italy}

\date{February 15, 2007}

\begin{abstract}
A technique inspired on quantum annealing is proposed in order to
obtain the classical ground state of a spin-glass by tracking the full
wavefunction of a given system within the subspace of matrix product
states (MPS), using the density matrix renormalization group
(DMRG). The technique is exemplified within the problem of obtention
of the classical ground state of an Ising spin glass on ladder
geometries. Its performance is evaluated and related to the
entanglement entropy.
\end{abstract}


\maketitle

\section{Introduction}

Global optimization is one of the most challenging numerical
tasks. Simulated thermal annealing (STA) has been, for more than
twenty years, one of the most popular general purpose tools. Its
strength relies on its ability to escape metastable minima via thermal
fluctuations. Simulated quantum annealing (SQA)
\cite{Das_Chakrabarti:book} is a more recent algorithm which takes
profit from {\em quantum} fluctuations for the same purpose. Both
methods have a clear physical motivation: thermal annealing has been
carried out by metallurgists for millenia. On the other hand, the
first evidence of the superiority of quantum over thermal fluctuations
in order to find the global minimum of a real physical system was only
obtained in 1999, in the context of quantum spin glasses
\cite{Brooke_Sci99}.

Numerical implementations of SQA have relied heavily on the {\em
quantum-classical} analogy described by the path integral Monte-Carlo
method (PIMC). Standard simulated thermal annealing is applied to an
enlarged system, which consists of several {\em replicas} of the
original system in interaction. Slightly different versions of the
method have succeeded in the obtention of the ground state of
classical spin glasses \cite{Kadowaki_PRE98,Santoro_SCI02}, atomic
clusters \cite{Lee_JPC00,Gregor_CPL05}, traveling salesman problem
\cite{Martonak_PRE04}, kinetically constrained systems
\cite{Das_PRE05} and small protein-folding problems \cite{Lee_JPC00},
among others. SQA has proved to work less efficiently than STA in
other problems, such as 3-SAT \cite{Battaglia_PRE05} and some
benchmark 1D potentials \cite{Stella_PRB05}.

Despite the success of the PIMC-SQA, there are reasons to look for
different implementations \cite{Battaglia:chapter}. First of all, PIMC
simulations must be carried out at finite temperature, with lower
temperatures requiring a larger number or replicas and, therefore,
higher computational cost. Also, some quantum systems suffer from the
sign problem, or from problems with the Trotter break-up. Sampling
difficulties are always a risk for annealing schemes, and ensuring
ergodicity is often a highly non-trivial task.

Advancing in this line, Green's function Monte-Carlo technique was
attempted by Stella and Santoro \cite{Stella_X06}. Its main
disadvantage is the necessity of good trial variational
wavefunctions. Real time evolution of the full wavefunction has been
implemented by Suzuki and Okada \cite{Suzuki:chapter}, making use of
both exact methods and the time-dependent density matrix
renormalization group (DMRG) algorithm \cite{White_PRL04}. For systems
where DMRG is efficient, loss of adiabaticity in the form of
Landau-Zener level crossings is the main problem of this approach.

The proposal of this work is to perform the annealing on the full
wavefunction, thus overcoming most of the difficulties associated with
PIMC. But, as opposed to the Suzuki-Okada approach, a real time
evolution is not needed either. Instead, the ground state of the full
system is computed exactly for a high intensity of the quantum
fluctuations. As these fluctuations are decreased, the ground state is
subsequently updated, by making the necessary small modifications of
the previous ground state. The full wavefunction of the ground state
is stored in the form of a matrix product state (MPS), and is computed
variationally using the DMRG.

As a benchmark case-study we have selected the obtention of the
classical ground state of an Ising spin-glass with couplings uniformly
distributed in $[-1,1]$. The quantum fluctuations are provided by a
transverse magnetic field. We have chosen the {\em ladders} for the
underlying topology, since it is the simplest case which presents
genuine frustration while retaining the quasi-1D character which is
required for the DMRG to attain its maximum efficiency. Albeit the
problem is in class P, we will show it to be complex enough to be
considered non-trivial.

This article is organized as follows. Section \ref{model} introduces
our model hamiltonian, summarizing its general features as applied to
our featured topology. It is followed, in section \ref{qwa}, by a
detailed description of the proposed quantum wavefunction annealing
(QWA) method. In section \ref{results} we analyze the results of the
numerical experiments we have conducted. We conclude, in section
\ref{discussion}, with a discussion of the advantages and limits of
the method, and its possible extensions.

%%%%%%%%%%%%%%%%%%%%%%%%%%%%%%%%%%%%%%%%%%%%%%%%%%%%%%%%%%%%%

\section{Random Ising model in a Transverse Field on Ladders}
\label{model}

Our model hamiltonian is the random Ising model in a transverse field
(RITF), which we define on an arbitrary graph ${\cal G}$ with $N$
sites \cite{Chakrabarti_Dutta_Sen:book}:

\begin{equation}
H=-\sum_{\<i,j\>} J_{ij} \sigma^z_i \sigma^z_j 
- \Gamma \sum_i \sigma^x_i 
- \sum_i h^z_i\sigma^z_i
\label{ritf}
\end{equation}

\nind where $\<i,j\>$ denotes nearest neighbours in ${\cal G}$, the
$J_{i,j}$ are independent random variables uniformly distributed in
$[-1,1]$, $\Gamma$ is the transverse field and $\sigma^x_i$ and
$\sigma^z_i$ are Pauli matrices at site $i$. We will work on the basis
of eigenstates of $\sigma^z$. Unless explicitly stated, we will assume
the longitudinal fields $h^z_i$ to be zero. The first term in
eq. \ref{ritf} may be considered as a potential energy, and the second
as a kinetic term: the magnitude of $\Gamma$ measures the intensity of
the quantum fluctuations. For $\Gamma=0$, the ground state is composed
of only two states, related by a trivial $+\Leftrightarrow -$
symmetry, which we term the classical ground states (CGS) of the spin
glass. As $\Gamma\to\infty$, on the other hand, all spins should be
pointing in the $X$-direction, therefore obtaining the following
ground state:

\begin{equation}
\ket|X>=\otimes_{i=1}^N \frac{1}{\sqrt{2}} \( \ket|+>_i + \ket|->_i \) 
\label{paramagnetic_state}
\end{equation}

\nind All the wavefunction components of the ground state take the
same value in our basis for $\Gamma\to\infty$.

The behaviour of this system at $T=0$ is rather well known in 1D
\cite{Fisher_PRB95,Fisher_PRB98} thanks to an insightful RG analysis,
in 2D \cite{Rieger_PRL94} and 3D \cite{Guo_PRL94} using quantum
Monte-Carlo, and in random graphs with fixed connectivity
\cite{Laguna_X06} making use of the DMRG. In all these cases, the
system presents a quantum spin-glass transition (QSGT) at a finite
value of $\Gamma=\Gamma_c$, where the energy gap vanishes. Above that
value, the system is said to be in a {\em quantum paramagnetic}
regime, whilst below it behaves as a {\em quantum spin-glass}.

The problem of finding the global CGS for equation \ref{ritf} at
$\Gamma=0$ is known to be in class P for $D=1$ and for $D=2$ in the
absence of longitudinal magnetic field $h^z$. In the 2D case with
$h^z\neq 0$, the 3D case \cite{Barahona_JPA82} or in random graphs
with fixed connectivity \cite{Liers_PRB03}, the problem is
NP-complete.
\vspace{5mm}

\begin{figure}
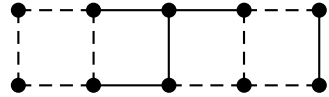

\psset{unit=1mm}
\def\la{\psline[linestyle=dashed]}
\def\lb{\psline}
\rput(-25,-10){
\multirput(10,0){5}{\pscircle*(0,0){1}}
\rput(0,-10){\multirput(10,0){5}{\pscircle*(0,0){1}}}
\la(0,0)(10,0)\la(0,0)(0,-10)
\lb(10,0)(20,0)\la(10,0)(10,-10)
\lb(20,0)(30,0)\lb(20,0)(20,-10)
\la(30,0)(40,0)\la(30,0)(30,-10)
\lb(40,0)(40,-10)
\la(0,-10)(10,-10)
\lb(10,-10)(20,-10)
\la(20,-10)(30,-10)
\la(30,-10)(40,-10)}
\vspace{25mm}
\caption{\label{ladder:fig}A rectangular ladder of dimension $5\times
2$. Coupling constants $J_{ij}$ are associated to links of the
graph. In the example of the figure, dashed (continuous) lines
represent negative-AFM (positive-FM) links, and the system is
frustrated.}
\end{figure}

{\em Rectangular ladders} constitute the simpest topology in which the
system described by equation \ref{ritf} presents genuine
frustration. Their size is characterized by two numbers: $L\times w$,
where $L$ is the length and $w$ is their width, or number of
legs. Figure \ref{ladder:fig} shows a specimen with size $5\times
2$. The classical spin-glass with $\pm J$ couplings on these ladders
has been analyzed by a number of authors
\cite{Mattis_PRL99,Uda_PRB05}. We will now discuss the complexity of
the energy landscape for $\Gamma=0$ and a few basic characteristics of
the nature of the QSGT.

\begin{figure}
\epsfig{file=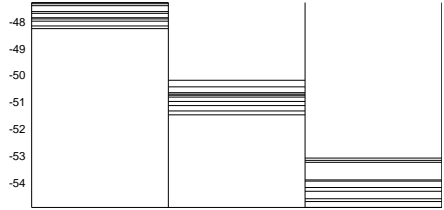,width=6.0cm,angle=0,clip=}
\caption{\label{sta_ladder:fig}Some local minima of the classical
  hamiltonian for three $40\times 2$ samples, obtained with a
  non-robust STA in order to sample the low energy configurations.}
\end{figure}

Minimization of the classical hamiltonian ($\Gamma=0$) can be done
using STA or PIMC-SQA. For the system sizes under consideration, both
methods yield robust estimates for the CGS energy following the
schemes described by Santoro and coworkers\cite{Santoro_SCI02}. In
order to analyse the complexity of the energy landscape, we relax the
parameters until the system becomes non-robust, i.e.: it yields
different results in different runs. At this point, the annealing
processes provide us a series of sample configurations which
constitute local minima of the energy. With this purpose, we have
applied a STA algorithm with a multiplicative scheme for $\beta$,
i.e.: $\beta\to r \beta$, from $\beta_0=0.1$ to $\beta_{max}=10^6$,
with $r=1+10^{-5}$ and $10^4$ steps per temperature. The results for
three samples of a $40\times 2$ ladder are shown in figure
\ref{sta_ladder:fig}. Each column contains the energy values obtained
after the procedure was repeated 20 times, providing several different
local energy minima (about 10). In all the cases shown, the lowest
energy corresponds to the CGS. This result points to a complex energy
landscape for the spin-glass ladders.

The energy per spin of the CGS seems to converge, in the case of the
2-legged ladders, to $\epsilon^{(2)} \approx -0.64$. In absence of
frustration, this value would be $-0.75$. In the case of the ladder
with 4 legs, it converges to $\epsilon^{(4)} \approx -0.71$, with the
unfrustrated value being $-0.875$.
\vspace{5mm}

\begin{figure}
\epsfig{file=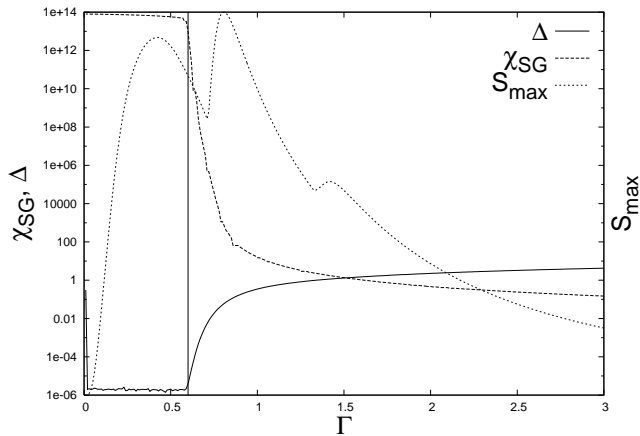,width=6.0cm,angle=270,clip=}
\caption{\label{qsgt_ladder:fig}Behavior of the energy gap $\Delta E$,
  the maximum block entropy $S_{max}$ and the spin-glass
  susceptibility $\chi_{SG}$ for a sample spin-glass ladder of
  dimension $40\times 2$, as a function of $\Gamma$. The QSGT is
  marked very clearly by both $\chi_{SG}$ and $\Delta E$. On the other
  hand, $S_{max}$ seems to have a more erratic behavior.}
\end{figure}

At $T=0$, the system described by eq. \ref{ritf} presents a quantum
spin-glass transition (QSGT) at a finite value of
$\Gamma=\Gamma_c$. The finite-size DMRG algorithm, suitably adapted
for our case\cite{Laguna_X06}, can be used to characterize this
transition. The behavior of some relevant observables has been traced
in figure \ref{qsgt_ladder:fig}: (a) the energy gap $\Delta E$, (b)
the maximum block entropy $S_{max}$, given by

\begin{equation}
S_{max}\equiv -\hbox{Tr} \rho_L \ln(\rho_L)
\label{entropy}
\end{equation}

where $\rho_L$ is the reduced density matrix for the left half of the
ladder, and (c) the spin glass susceptibility $\chi_{SG}$, as defined
by the following formula:

\begin{equation} 
\chi_{SG}\equiv \frac{1}{N} \sum_{i,j} \lim_{h^z_j\to 0}
\( \frac{\<\sigma^z_i\>}{h^z_j} \) ^2 \;,
\label{xsg}
\end{equation}

\noindent i.e.: a small longitudinal magnetic field $h^z_j$, applied
at site $j$, generates a magnetization response on each site $i$,
which is measured (and squared, so as to disregard its sign); the
results are summed over all sites $i$ and averaged over all sites $j$.
The value of $\Gamma_c$ changes from sample to sample. Within a single
sample, $\Delta E$ vanishes and $\chi_{SG}$ diverges at the same value
of $\Gamma_c$, as it is shown in figure \ref{qsgt_ladder:fig}. For
that sample, $\Gamma_c=0.6$. The entropy is seen to present a more
complex behavior.

\begin{figure}
\epsfig{file=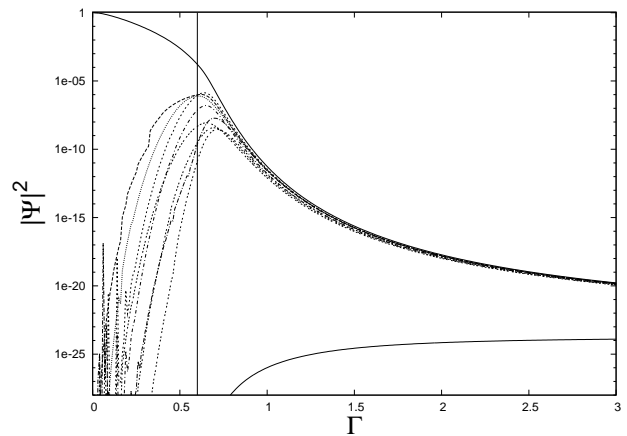,width=6.0cm,angle=270,clip=}
\caption{\label{wfc:fig}Wavefunction components for some selected
  configurations, as a function of $\Gamma$, for a sample ladder with
  size $40\times 2$. All the configurations but one are local minima
  of the classical energy function, obtained using a non-robust
  STA. The transition point is marked with an vertical line.}
\end{figure}

Some insight can be gained by tracing a few individual components of
the full wavefunction, which is possible within the DMRG framework
\cite{Laguna_X06}. Figure \ref{wfc:fig} shows that, well within the
paramagnetic regime, all wavefunction components take the same
value. As we decrease $\Gamma$, the configurations with low energy
start to increase its weight in the ground state wavefunction, while
the configurations with high energy start to decrease. At the critical
point, all the configurations have started their decay, except the one
with the lowest energy.

More theoretical and analytical work is needed in order to fully
characterize this QSGT. Some comments about it are provided in section
\ref{discussion}. This preliminary analysis can be summarized as: (a)
the energy landscape of the classical problem is complex, and (b) there
is a quantum phase transition at a finite value of $\Gamma$.

%%%%%%%%%%%%%%%%%%%%%%%%%%%%%%%%%%%%%%%%%%%%%%%%%%%%%%%%%%%%%%%%%

\section{Quantum Wavefunction Annealing of Spin Glass Systems}
\label{qwa}

Quantum wavefunction annealing (QWA) is an implementation of simulated
quantum annealing (SQA) which does not require real time evolution,
configurations sampling or Trotter splitting. It makes extensive use
of matrix product states (MPS) \cite{Fannes_CMP92,Roman_JPA98} and the
density matrix renormalization group (DMRG) \cite{White_PRL92}. For
the benefit of the readers which are not acquainted with those
techniques, the first paragraph of this section describes a simpler
formulation based on the Lanczos algorithm, but which can only be used
for very small lattices.

The Lanczos diagonalization method improves its performance
dramatically if a good seed to the real ground state of the system is
provided \cite{Golub:book}. The hamiltonian given by equation
\ref{ritf} has a simple behaviour for $\Gamma\to\infty$: its ground
state is known to be given by the state $\ket|X>$ defined in equation
\ref{paramagnetic_state}. Therefore, starting with such a state as a
seed, it is fairly easy for the Lanczos algorithm to obtain the exact
ground state for a finite but high value $\Gamma_0$. This value may be
decreased at small finite steps, $\Gamma\to\Gamma-\Delta\Gamma$, and
the ground state for the previous value of $\Gamma$ can be employed as
the seed for the computation of the ground state at the new value. In
quantum annealing, the adiabatic theorem is needed in order to prove
that convergence to the CGS is sure in the limit $\Delta\Gamma\to
0$. In our case, a weaker result is enough: convergence is sure as
long as there is a finite overlap between the ground states at any two
consecutive values of $\Gamma$.

The obvious drawback of the previous algorithm is the size of the
Hilbert space, $\dim({\cal H})=2^N$. Using the plain Lanczos
algorithm, all the wavefunction components should be stored and acted
upon. A solution is to choose a low-dimensional subspace which is
known to contain the ground state of the hamiltonian given by equation
\ref{ritf} for all values of $\Gamma$. Matrix product states (MPS) can
provide such a subspace in some cases. These states may be written
down as:

\begin{equation}
\ket|\Psi>=\sum_i \hbox{tr}\( A^{s_1} A^{s_2} \cdots A^{s_N} \)
\ket|s_1,s_2,\cdots,s_N> 
\label{mps}
\end{equation}

\nind where $\ket|s_1,s_2,\cdots,s_N>$ are the eigenstates of
$\sigma^z_i$ and the $A^{s_i}$ are $2N$ matrices of dimension $m\times
m$, whose entries are the variational parameters of our Ansatz. The
dimension of the MPS subspace is, therefore, bounded by $2Nm^2$. Both
the computational cost and the accuracy of the method depend strongly
on the dimension $m$ of the matrices, whose physical meaning is the
following: for all possible left-right splittings of the system, the
ground state is approximated retaining $m$ states to represent the
left part and other $m$ states for the right part. For any possible
state of the system, its representation as an MPS becomes exact for
$m$ large enough. Of course, the dimensions $m$ can be made local,
$m_i$, if necessary.

The DMRG is a variational method within the subspace of the MPS
\cite{Rommer_PRB97}. It profits from the use of a density matrix in
order to select the $m$ states for the left and right blocks which fit
best to our current approximation of the global ground state. The
neglected eigenvalues of the density matrix provide a way to monitor
the accuracy of the method. In all our applications, the tolerance to
the total {\em neglected probabilities} is fixed beforehand to a
certain value $\eta$, and the number of retained states $m(\eta)$, is
adapted in consonance.

The MPS represent {\em faithfully} ground states of local 1D
hamiltonians \cite{Verstraete_PRB06}. In other cases, at this stage,
only educated experience can decide whether they are appropriate or
not. They are specially suited to 1D and quasi-1D problems
(e.g. ladders, trees), although natural extensions to higher
dimensions are in active development
\cite{Nishino_PTP01,Verstraete_X04}. A crucial tool is von Neumann's
{\em block entropy}, $S=-\hbox{tr}(\rho \log_2 \rho)$. In crude terms,
$m$ should scale as $\exp(S)$ in order to obtain an accurate DMRG
method. In 1D, the entropy $S$ is known to be bounded for non-critical
systems and to scale as $\log(N)$ for a critical one
\cite{Vidal_PRL02}. For higher dimensions, the {\em area law} predicts
the entropy to scale as $L^{d-1}$ out of criticality, where $L$ is a
typical dimension of the system \cite{Sredniki_PRL93}.

As in the Lanczos case, the DMRG can benefit from a good {\em seed}
when searching the ground state of a hamiltonian through the use of
the {\em wavefunction transformations} \cite{White_PRL96}, which allow
to use the solution of an RG-step as an initial step for the next
one. Having found the ground state of a certain hamiltonian $H$, a few
finite-size sweeps will {\em adapt} it to become the ground state of a
slightly modified hamiltonian $H'$. In normal cases, this procedure is
not needed, since the number of sweeps required for convergence to the
ground state for any hamiltonian is small. It may become very useful
when (a) it is known that the ground state of a certain hamiltonian
{\em can} be represented as a MPS with low $m$ but (b) the probability
of the DMRG getting stuck at an excited state is very high. This is
the case for the ground state of \ref{ritf} in the quantum spin-glass
phase\cite{Laguna_X06}.

Our proposed QWA algorithm is, therefore, the following one:

\begin{description}

\item[{\bf (a)}] The ground state of the system \ref{ritf} is obtained
for very high $\Gamma$.

\item[{\bf (b)}] The transverse field $\Gamma$ is decreased 
$\Gamma\to \Gamma-\Delta\Gamma$.

\item[{\bf (c)}] A few finite-size sweeps of the DMRG are carried out,
which adapt the ground state to the current value of $\Gamma$.

\item[{\bf (d)}] Go to (b) if $\Gamma$ is not yet zero.

\item[{\bf (e)}] Measure the energy and classical ground state.

\end{description}

This approach is {\em deterministic}, i.e.: not limited by sampling
problems. It works directly with the quantum hamiltonian, therefore
does not require any Trotter break-up, and works directly at
$T=0$. Since it does not simulate real time evolution, it is not prone
to Landau-Zener level-crossings. Loss of adiabaticity is not,
therefore, a crucial issue. As it was stated, it is enough to ensure
that the overlap between the ground states at consecutive values of
$\Gamma$ is finite. A high value of $\Delta\Gamma$ is allowed far away
from the quantum phase transition, while a more reduced value will be
taken near it. Our precise adaptive reduction schedule is described in
the following section.

On the negative side, it is based on a method which is specially
designed for 1D systems. We will say more about this in the
conclusions. Its main weakness stems from the inability of the MPS to
represent faithfully the ground state for all values of $\Gamma$ for
higher dimension. The efficiency and practical issues of the
implementation are analyzed in the next section.

%%%%%%%%%%%%%%%%%%%%%%%%%%%%%%%%%%%%%%%%%%%%%%%%%%%%%%%%%%%%%%%%%

\section{Numerical Experiments}\label{results}

In order to study the applicability of QWA, we have generated random
samples of ladders with $w=2$ and $w=4$ legs, and lengths ranging from
$L=20$ to $L=160$. We have also considered random graphs of fixed
connectivity $K=3$ and, as a check, we have also analysed linear
chains (although the optimization problem is trivial in that
case). The annealing scheme has always been the same: starting with
$\Gamma_0=3$ and using an adaptative reduction,
$\Gamma\to\Gamma-\Delta\Gamma$ with

\begin{equation}
\Delta\Gamma=\min(0.5,0.1/S)
\label{annealing_scheme}
\end{equation}

\noindent where $S$ is the maximum block entropy for the previous
value of $\Gamma$. This way we ensure that, near the QSGT --when the
entropy is higher-- the steps are shorter. We insist in this fact: the
adiabatic theorem and Landau-Zener level crossings are not the
limitations of this method. Therefore, a decrease in the size of the
steps does not necessarily increase the probability of success.

The last annealing step was always taken with $\Gamma_{min}=0.01$,
a value which is low enough for practical purposes. The obtention of
the minimum energy configuration was carried out measuring the
$z$-component of the spin polarization of all sites after convergence
for $\Gamma=\Gamma_{min}$. Of course, due to the obvious symmetry
$+\Leftrightarrow -$, the expectation value for each of these
components is zero. We have used a common procedure in quantum
spin-glass calculations: the insertion of a very small longitudinal
magnetic field $h^z=10^{-6}$ in a single site, selected at random, in
order to break the symmetry. In the QSG phase, because of the
divergence of the spin-glass susceptibility discussed above, an
infinitesimal localized magnetic field polarizes the full sample.

Table \ref{results:table} provides the basic set of results. For each
geometry we have selected 20 samples and performed the QWA algorithm
on them. We have also chosen different values of the neglected
probabilities tolerance, $\eta$. For each run, QWA is said to obtain a
success if its minimum energy is equal to the value obtained by STA
and PIMC-SQA. QWA has had success, in all the attempted geometries,
whenever $\eta=10^{-8}$. When the tolerance was decreased, the CGS was
missed with higher probability, although the method can be seen to be
rather robust for spin-glass ladders, since the tolerance must be
raised up to $10^{-3}$ in order to decrease the probability of success
to $50\%$.

\begin{table}
\begin{ruledtabular}
\begin{tabular}{lccc}
Geom. & $\eta$ & Success & Time \cr
$20$          & $10^{-8}$ & 100\% & $1.2\pm 0.2$   \cr
$40$          & $10^{-8}$ & 100\% & $3.9\pm 0.8$   \cr
$80$          & $10^{-8}$ & 100\% & $10\pm 2$      \cr
$160$         & $10^{-8}$ & 100\% & $21\pm 4$      \cr
$320$         & $10^{-8}$ & 100\% & $56\pm 10$     \cr
\hline
\vspace{1mm}
$20\times 2$  & $10^{-8}$ & 100\% & $17\pm 8$      \cr
$40\times 2$  & $10^{-8}$ & 100\% & $60\pm 15$     \cr
$80\times 2$  & $10^{-8}$ & 100\% & $240\pm 30$    \cr
$160\times 2$ & $10^{-8}$ & 100\% & $600 \pm 110$  \cr
\hline
\vspace{1mm}
$40\times 2$  & $10^{-5}$ & 80\%  & $26\pm 3$      \cr
$40\times 2$  & $10^{-3}$ & 50\%  & $13\pm 1$      \cr
\hline
\vspace{1mm}
$20\times 4$  & $10^{-8}$ & 100\% & $1600\pm 600$  \cr
$40\times 4$  & $10^{-8}$ & 100\% & $6800\pm 2000$ \cr
$80\times 4$  & $10^{-8}$ & 100\% & $14000\pm 3000$\cr
$160\times 4$ & $10^{-8}$ & 100\% & $27000\pm 3000$\cr
\hline
\vspace{1mm}
RG-$20$       & $10^{-8}$ & 100\% & $190\pm 150$   \cr
RG-$100$      & $10^{-6}$ & 45\%  & $15000\pm 7000$\cr
\end{tabular}
\end{ruledtabular}
\caption{\label{results:table}Numerical results for the QWA
method. The first column states the geometry of the system under
study: $L$ if it is a linear chain, $L\times w$ if it is a ladder, and
RG-$N$ if it is a random graph with connectivity $K=3$. The second
gives the tolerance for neglected probabilities in the DMRG,
$\eta$. The third provides the percentage of success. The fourth
column provides the average time (in seconds) for the QWA of a single
sample.}
\end{table}

We have also tried to check the method with a different graph
topology: random graphs with fixed connectivity, $K=3$. For $N=20$
sites, employing a tolerance of $\eta=10^{-8}$, the method provides
again 100\% of success. If the size is increased, that tolerance can
not be set, since the number of states $m$ to be kept becomes
prohibitive. Therefore, we have carried out the experiments with
$\eta=10^{-6}$ for a $N=100$ sample, and the probability of success
reduces to $45\%$.

The time for the QWA algorithm scales as a power law of the system
size: $T\approx L^{\alpha}$. For linear chains, using the results from
table \ref{results:table}, $\alpha\approx 1.3\pm 0.1$. For 2-legged
ladders, $\alpha\approx 2\pm 0.1$. In the case of the 4-legged
ladders, the power law fit has a higher errorbar, with a surprising
exponent of $\alpha\approx 1.5\pm 0.2$.

\subsection{QWA and block entropy}

This polynomial growth can be theoretically explained. For quasi-1D
systems at criticality, the von Neumann entropy grows as a logarithm
of the system size\cite{Vidal_PRL02}, $S_c(L)\approx a\log(L) +
b$. The value of $a$ has been related, in some random systems, to the
central charge of the associated conformal field
theory\cite{Refael_PRL04}. On the other hand, the maximum number of
retained states scales as the exponential of the entropy, $m\approx
\exp(S)$, and the time for a DMRG sweep scales as $T\approx Lm^2$. The
most expensive DMRG sweep for a QWA simulation takes place at the
critical point, and therefore we may expect $T\approx
L\exp(2S_c)\approx L^{2a+1}$. Thus, the theoretical prediction is
$\alpha\simeq 2a+1$.

In the linear chain case, the entropy at criticality was predicted by
Refael and Moore \cite{Refael_PRL04} to have a coefficient
$a=\ln(2)/6\approx 0.11$, in agreement with our own numerical
measurements. Therefore, the theoretical prediction for the $\alpha$
exponent is $\approx 1.22$, which is not far from the $1.3\pm 0.1$
obtained numerically. In the case of the 2-legged ladder, there is no
theoretical estimate, but our own numerical simulations provide a
value $a\approx 0.55\pm 0.1$, thus giving an estimate for $\alpha$
around $2.1$, in agreement with the numerically observed value
$\alpha\approx 2 \pm 0.1$.

%%%%%%%%%%%%%%%%%%%%%%%%%%%%%%%%%%%%%%%%%%%%%%%%%%%%%%%%%%%%%%%%%%%

\section{Conclusions and Further Work}\label{discussion}

In this paper we have introduced a method for the obtention of the
minimum energy configuration of a spin-glass, based on the annealing
of the full wavefunction represented as a matrix product state (MPS)
using the density matrix renormalization group (DMRG). The method has
been termed {\em quantum wavefunction annealing} (QWA), and we have
assessed and quantified its efficiency, by comparing its results with
those provided by other robust methods. For spin-glass ladders of 2
and 4 legs with lengths ranging from $20$ to $160$, the method has
always provided the optimum solution when the probability tolerance is
set to $\eta=10^{-8}$. The running time scales as $L^2$ for 2-legged
ladders, and, surprisingly, as $L^{1.5}$ for 4-legged ones. This
anomaly requires further clarification, perhaps in the line of thought
of Ferraro et al \cite{Ferraro_X07}. For random graphs with fixed
connectivity $K=3$, on the other hand, the tolerance has to be
increased in practice to $10^{-6}$, and the probability of success
falls to $45\%$ for $N=100$.

The most crucial parameter which determines the success of the method
is the probability tolerance $\eta$. Whenever we were able to set
$\eta=10^{-8}$, for whichever topology, the system always attained the
optimum configuration. Unfortunately, for topologies which are not
quasi-1D, the number of retained states in DMRG, $m$, grows very fast
when the tolerance is decreased. Other parameters, such as the
annealing velocity, do not have such a direct relevance to the quality
of the results. Loss of adiabaticity is not the main issue for this
method: a finite overlap between the ground state wavefunctions at
consecutive annealing steps is enough to ensure convergence, as long
as the number of retained states $m$ is large enough for our Ansatz
wavefunction to represent both of them faithfully.

Therefore, it seems reasonable to think that QWA can be converted into
a method to find the optimum configuration of all quasi-1D systems in
polynomial time. This suggests that, in fact, no quasi-1D problem can
be NP-complete. On the other hand, for problems which are known to be
NP-complete, such as the random Ising spin-glass in a 2D lattice with
a longitudinal field, or on random graphs of fixed connectivity, the
time for an algorithm which is able to obtain the optimum with
certainty should scale exponentially with the size of the system.

This fact puts a limit to possible extensions of the QWA algorithm
described in this paper. The natural extension of the MPS are the {\em
tensor product states} (TPS) analyzed by Nishino and coworkers
\cite{Nishino_PTP01} or the pair-entangled particle states (PEPS)
\cite{Verstraete_X04} of Verstraete and Cirac. The computational power
of this type of states has been analyzed recently \cite{Schuch_X06},
pointing to the fact that any attempt to solve NP-complete problems
with an exact QWA algorithm using these states would require an
exponential time, which is compatible with the usually believed notion
that P $\neq$ NP.

The most promising lines for future work are, therefore, in the
development of heuristic algorithms, running in polynomial time, which
may give the absolute minimum energy state with high probability for
{\em some} problems. A possibility is the development of a QWA
algorithm using TPS (or PEPS) with a fixed number of retained states
$m$. Another possibility is to use the DMRG approach to
non-equilibrium classical problems \cite{Degenhard_MMS04} to develop a
classical wavefunction annealing algorithm, which would solve the
Fokker-Planck equation associated to simulated thermal annealing.

\begin{acknowledgments}
The author acknowledges G.~Santoro, A.~Trombettoni,
M.A.~Mart\'{\i}n-Delgado and G.~Sierra for very useful
discussions. 
\end{acknowledgments}

%\section*{References}

% Comment if you want to insert a .bbl file directly below
%\bibliography{Disorder}

\end{document}